\documentclass[twocolumn,10pt, nofootinbib]{revtex4-1}
\usepackage{times}
\usepackage{amsmath}
\usepackage{amsfonts}
\usepackage{graphics,graphicx}
\usepackage{color}
\usepackage{tabularx}
\usepackage[T1]{fontenc}
\usepackage{booktabs}

\newcommand{\citen}[1]{\onlinecite{#1}}
\newcommand{\overbar}[1]{\mkern 1.5mu\overline{\mkern-1.5mu#1\mkern-1.5mu}\mkern 1.5mu}

\newcommand{\sVRH}{{\substack{\scalebox{0.6}{VRH}}}}

\begin{document}  
\title{Predicting Superhard Materials via a Machine Learning Informed Evolutionary Structure Search}

\author{Patrick Avery}
  \affiliation{Department of Chemistry, State University of New York at Buffalo, Buffalo, NY 14260-3000, USA}
\author{Xiaoyu Wang}
  \affiliation{Department of Chemistry, State University of New York at Buffalo, Buffalo, NY 14260-3000, USA}
\author{Davide M. Proserpio}
  \affiliation{Dipartimento di Chimica, Universit\`a degli Studi di Milano, Via Golgi 19, 20133 Milano, Italy}
  \affiliation{Samara Center for Theoretical Materials Science (SCTMS), Samara State Technical University,
Molodogvardeyskaya St.\ 244, Samara 443100, Russia}
\author{Cormac Toher}
  \affiliation{Department of Mechanical Engineering and Materials Science, Duke University, Durham, NC, 27708, USA}
  \affiliation{Center for Materials Genomics, Department of Mechanical Engineering and Materials Science, Duke University, Durham, North Carolina 27708, USA}
\author{Corey Oses}
  \affiliation{Department of Mechanical Engineering and Materials Science, Duke University, Durham, NC, 27708, USA}
  \author{Eric Gossett}
  \affiliation{Department of Mechanical Engineering and Materials Science, Duke University, Durham, NC, 27708, USA}
  \author{Stefano Curtarolo} \email{stefano@duke.edu}
  \affiliation{Department of Mechanical Engineering and Materials Science, Duke University, Durham, NC, 27708, USA}
  \affiliation{Center for Materials Genomics, Department of Mechanical Engineering and Materials Science, Duke University, Durham, North Carolina 27708, USA}
  \affiliation{Fritz-Haber-Institut der Max-Planck-Gesellschaft, 14195, Berlin-Dahlem, Germany}
\author{Eva Zurek}\email{ezurek@buffalo.edu}
  \affiliation{Department of Chemistry, State University of New York at Buffalo, Buffalo, NY 14260-3000, USA}

\date{\today}

\begin{abstract}

Good agreement was found between experimental Vickers hardnesses, $H_\text{v}$, of a wide range of materials and those calculated by three macroscopic hardness models that employ the shear and/or bulk moduli obtained from: (i) first principles via AFLOW-AEL (AFLOW Automatic Elastic Library), and (ii) a machine learning (ML) model trained on materials within the AFLOW repository. Because $H_\text{v}^\text{ML} $ values can be quickly estimated, they can be used in conjunction with an evolutionary search to predict stable, superhard materials. This methodology is implemented in the \textsc{XtalOpt} evolutionary algorithm. Each crystal is minimized to the nearest local minimum, and its Vickers hardness is computed via a linear relationship with the shear modulus discovered by Teter. Both the energy/enthalpy and $H_\text{v, Teter}^{\text{ML}}$ are employed to determine a structure's fitness. This implementation is applied towards the carbon system, and 43 new superhard phases are found. A topological analysis reveals that phases estimated to be slightly harder than diamond contain a substantial fraction of diamond and/or lonsdaleite.
\end{abstract}

\maketitle

\renewcommand{\thefootnote}{\roman{footnote}}

\section{\label{sec:Intro}Introduction}
Superhard materials are important in a wide variety of applications including cutting and polishing, or as abrasives and coatings. 
They typically contain light elements such as B, C, N, and O, which can form short and strong covalent bonds, and they may have complex potential energy surfaces (PES) with numerous low lying minima. Because the main atomic constituents of superhard materials all have similar masses, it is sometimes impossible to determine their crystal structures from X-ray diffraction patterns alone. As a result, first-principles calculations have been instrumental in uncovering the structures of a number of superhard phases including $R3m$-BC$_2$N \cite{Li:2009cb2n}, the $M$-carbon phase formed upon cold compression of graphite \cite{mao2003bonding,Wang:2012a,Li:2009a}, $Pnnm$-CN \cite{Stavrou:2016a}, and cubic BC$_3$ \cite{Zhang:2015a}. 

Many superhard materials are metastable at ambient pressure/temperature conditions, and the synthesis procedure can have an impact on the product that is made. Therefore, a synergistic feedback loop between theory and experiment is required for the rational design of superhard phases with specific properties. For example, recently an entropy-forming ability descriptor was developed and used to predict novel carbides containing five metals with a high hardness that were subsequently synthesized \cite{sarker:2018a}. Another example is the development of machine learning (ML) models for the elastic moduli, which were subsequently employed to screen hundreds of thousands of compounds found in a crystal structure database. The most promising superhard ternaries, Re$_{0.5}$W$_{0.5}$C and ReWC$_{0.8}$, were subsequently synthesized and experiments confirmed they were superhard at low load \cite{Tehrani:2018a}.

The \textit{a priori} prediction of superhard materials is a global optimization problem that requires maximizing the computed hardness in the space of the low-lying local minima on the PES.  Microscopic models employing quantities that can be readily obtained such as the geometry of the crystal, atomic connectivity, valence electron density, and electronegativity or ionicity of the constituent atoms \cite{Gao:2003a,Simunek:2006a,Li:2008a,Zhang:2013a,Lyakhov:2011a} can be used to estimate a structure's hardness. Recently, the well-known crystal structure prediction (CSP) algorithms CALYPSO \cite{Zhang:2013a} and USPEX \cite{Lyakhov:2011a} have been extended towards the prediction of superhard phases. The structures generated in the course of a CSP search with these programs are typically optimized using first-principles calculations, and their hardnesses are calculated using modified versions of either the  \v{S}im$\mathring{\text{u}}$nek-Vack\'a$\check{\text{r}}$ (SV) \cite{Simunek:2006a} or the Li \textit{et al.}\ \cite{Li:2008a}  methods, respectively. The main deficiency of the original Li and SV equations is that they predict unreasonably high hardness values for crystals with non-bonding interactions that are necessary for maintaining the three dimensional structure of the system, such as the van der Waals forces between different layers of graphene in graphite. The hardness models developed in CALYPSO and USPEX overcome these limitations by using chemical graph theory to determine which atoms are bonded to each other.

The hardness of a material can also be estimated given its elastic properties, such as the bulk and shear moduli \cite{Teter:1998a,Chen:2011a,Tian:2012a}. One of the advantages of these macroscopic hardness models is that they do not depend explicitly upon quantities that are ambiguous or hard to define, such as the atomic radii, bond electronegativity, or the bond strength. However, because elastic properties are expensive to calculate from first-principles, they have so far not been coupled with CSP searches where the hardness of hundreds, if not thousands, of structures would need to be obtained within a single run. Models that require the computed elastic properties are typically only employed to determine the hardness of a handful of promising structures found within a search. 

Large materials databases that contain many measured and calculated observables \cite{Curtarolo:2012a,Saal:2013a,Landis:2012a,Bergerhoff:1983a,Jain:2013a,deJong:2015a} have allowed for the advent of ML models that have demonstrated predictive power for numerous properties, for example superconducting critical temperatures \cite{Stanev:2018a}, electronic band gaps \cite{Isayev:2017a,Pilania:2016a}, elastic properties \cite{Isayev:2017a}, and the melting temperatures of unary and binary component solids \cite{Seko:2014a}. AFLOW (Automatic FLOW) is an automatic framework for high-throughput materials discovery \cite{Curtarolo:2012a} that includes a materials database with over 2 million entries \cite{Curtarolo:2012b}. Many properties have been calculated for the materials in the AFLOW database including vibrational properties with the Automatic Phonon Library (APL) \cite{Curtarolo:2012a}, thermal properties with the Automatic GIBBS Library (AGL) \cite{Toher:2014a}, and thermomechanical properties such as the bulk and shear moduli with the Automatic Elasticity Library (AEL) \cite{Toher:2017a, deJong:2015a}. It is now possible to interact with the AFLOW database via a RESTful API \cite{Taylor:2014a} and the AFLUX materials search API \cite{Rose:2017a}.

Herein, we illustrate that the Vickers hardnesses, $H_\text{v}$, of a wide variety of crystalline materials predicted by using a macroscopic hardness model in conjunction with ML-derived bulk and shear moduli obtained via the RESTful API \cite{Gossett:2018a}  available on AFLOW are in excellent agreement with results obtained from first-principles calculations. Both are in good agreement with experiment. These developments make it possible to quickly calculate reasonable hardness values for a given crystal structure using ML-based elastic properties, and these hardness estimates can subsequently be employed to calculate an individual's fitness in a CSP algorithm designed for the prediction of superhard phases. This technique is implemented within the \textsc{XtalOpt} evolutionary algorithm (EA) \cite{Lonie:2011a,Zurek:2018j}, and is subsequently applied towards the carbon system to search for novel stable and superhard phases. Seventy-nine dynamically stable, low energy, distinct topologies with $H_\text{v}>$~40~GPa are found in our searches. Forty-three of the predicted structures have not been reported previously.

\section{Results and Discussion \label{sec:results}}

\subsection{Macroscopic Hardness Models Coupled with Machine Learning}
Because the chemical bonding within a crystalline lattice affects the bulk modulus, $B$, of the material, it has been proposed that $B$ can be a good indicator of hardness \cite{Liu:1989a,Sung:1996a}. While this is true for specific classes of materials, such as diamond-like semiconductors, it turns out that, although some exceptions exist, the shear modulus, $G$, is a much better predictor of hardness \cite{Teter:1998a,Gao:2010a,Brazhkin:2002a}. The linear correlation between hardness and $G$ was originally noted by Teter in 1998 \cite{Teter:1998a}, and in 2011, Chen \textit{et al.}\ used the geometric shape of the Vickers indenter to derive the correlation coefficient \cite{Chen:2011a}. We refer to the resulting equation for estimating hardness,
\begin{equation}\label{eqn:TeterModelVickersHardness}
H_\text{v,Teter} = 0.151G ,
\end{equation}
as the \emph{Teter model} to distinguish it from another model developed by Chen and co-workers in the same manuscript. 
Chen \textit{et al.}\ suggested that discrepancies between $H_\text{v,Teter}$
and the experimentally measured Vickers hardnesses resulted from neglecting the plastic deformations in Eq.\ \ref{eqn:TeterModelVickersHardness}. In order to better account for these effects, they proposed a new formula that employs the famous Pugh's modulus ratio, $k = G/B$, which correlates well with a material's brittleness.
The parameters in the modified equation, which we refer to as the \emph{Chen model}, were obtained by an empirical fit yielding
\begin{equation}\label{eqn:ChenModelVickersHardness}
H_\text{v,Chen} = 2(k^2 G)^{0.585}  - 3 .
\end{equation}
Eq.\ \ref{eqn:TeterModelVickersHardness} was found to work well for brittle materials with a large $k$. The Chen model, Eq.\ \ref{eqn:ChenModelVickersHardness}, yielded better estimates of the hardness in most cases, with one exception being crystals with a low $k$ where the bonding was primarily metallic, such as fcc Al.

In 2012 Tian and co-workers noted that the intercept term in Eq.\ \ref{eqn:ChenModelVickersHardness} did not have a physical basis, and would yield negative values for some materials such as KI and KCl \cite{Tian:2012a}. Therefore, they obtained a revised formula, which we refer to as the \emph{Tian model}, via refitting the original function proposed by Chen \textit{et al.}\ without the intercept term as
\begin{equation}\label{eqn:TianModelVickersHardness}
H_\text{v,Tian} = 0.92k^{1.137}G^{0.708} .
\end{equation}
The Vickers hardnesses estimated with the formulae of Chen and Tian are in good agreement with experimental measurements for many systems, but they tend to overestimate the hardness of ionic compounds and materials for which $H_\text{v}$ is less than \textasciitilde 5 GPa.

The two parameters in the above equations, $G$ and $B$, have been computed for over 5000 unique materials in the AFLOW database via AEL \cite{Curtarolo:2012b}. We employed their Voigt-Reuss-Hill average values, $G_\sVRH$, and $B_\sVRH$, to estimate the Vickers hardnesses of 64 systems, including many of those studied by Chen \cite{Chen:2011a} (data for the full set can be found in the SI; Table \ref{tab:Compare_hard} lists results only for those systems considered in Ref.\ \citen{Chen:2011a}). This set of materials spanned a wide range of hardness values ($H_\text{v}^\text{Exp}=0.2-96.0$~GPa), and included ionic and covalently bonded crystals, as well as intermetallics. It has been demonstrated that AFLOW-AEL calculates reasonable bulk and shear moduli \cite{Toher:2017a}, so it should not be a surprise that the Vickers hardnesses obtained using them are in good agreement with experiment, and the predicted values reflect the variation in the results of the models themselves. 
\begin{table*}[!ht]
\setlength{\tabcolsep}{8pt}
\renewcommand{\arraystretch}{1.2}
  \centering
  \caption{\label{tab:Compare_hard} A comparison of the Vickers hardness, $H_\text{v}$, for various materials computed via the Teter (Eq.\ \ref{eqn:TeterModelVickersHardness}), Chen (Eq.\ \ref{eqn:ChenModelVickersHardness}), and Tian (Eq.\ \ref{eqn:TianModelVickersHardness}) models using the Voigt-Reuss-Hill (VRH) averages of the bulk, $B$, and shear, $G$, moduli obtained from the Automatic Elasticity Library (AEL) and via a Machine Learning (ML) model trained on the AFLOW database, with experiment. $H_\text{v}$, $B$ and $G$ are given in units of GPa. The dataset used was taken from Ref.\ \citen{Chen:2011a}, which also provides the experimental values. A full table with results for all of the 64 structures used in the correlation analyses and the plots in Fig.\ \ref{fig:ML_AEL_EXP} is provided in the SI.}
\begin{tabular}{l|c|ccccc|ccccc}
\hline\hline
Name	&	$H_\text{v}^\text{Exp,}\footnote{Experimental data are taken from Ref.\ \citen{Chen:2011a}.}$	&	$G_{\sVRH}^\text{AEL}$	&	$B_{\sVRH}^\text{AEL}$	&	$H_\text{v,Teter}^\text{AEL}$	&	$H_\text{v,{Chen}}^\text{AEL}$	&	$H_\text{v,{Tian}}^\text{AEL}$	&	$G_{\sVRH}^\text{ML}$	&	$B_{\sVRH}^\text{ML}$	&	$H_\text{v,Teter}^\text{ML}$	&	$H_\text{v,Chen}^\text{ML}$	&	$H_\text{v,Tian}^\text{ML}$	\\
\hline							
Diamond	&	96.0	&	517.9	&	434.0	&	78.2	&	92.2	&	93.9	&	514.0	&	430.2	&	77.6	&	91.9	&	93.5	\\
BC$_2$N	&	76.0	&	411.6	&	372.5	&	62.2	&	73.1	&	73.1	&	363.6	&	354.0	&	54.9	&	62.0	&	61.6	\\
c-BN	&	66.0	&	380.3	&	372.4	&	57.4	&	63.2	&	63.2	&	386.8	&	332.7	&	58.4	&	74.9	&	74.2	\\
$\beta$-SiC	&	34.0	&	186.6	&	212.4	&	28.2	&	33.6	&	32.2	&	163.8	&	201.5	&	24.7	&	28.0	&	26.9	\\
SiO$_2$	&	33.0	&	201.2	&	270.1	&	30.4	&	28.5	&	28.1	&	187.2	&	237.8	&	28.3	&	29.3	&	28.5	\\
ReB$_2$	&	30.1	&	270.0	&	331.8	&	40.8	&	38.6	&	38.3	&	257.9	&	333.4	&	38.9	&	35.1	&	35.0	\\
WC	&	30.0	&	272.3	&	377.4	&	41.1	&	33.3	&	33.6	&	247.0	&	369.0	&	37.3	&	28.4	&	28.8	\\
VC	&	29.0	&	215.6	&	295.7	&	32.6	&	29.0	&	28.8	&	164.1	&	274.8	&	24.8	&	18.6	&	18.9	\\
ZrC	&	25.8	&	157.0	&	220.9	&	23.7	&	22.8	&	22.4	&	136.0	&	194.4	&	20.5	&	20.3	&	19.9	\\
TiC	&	24.7	&	177.8	&	238.8	&	26.8	&	26.3	&	25.8	&	145.9	&	220.8	&	22.0	&	19.7	&	19.6	\\
TiN	&	23.0	&	171.2	&	269.8	&	25.9	&	20.8	&	20.9	&	141.5	&	240.5	&	21.4	&	16.5	&	16.8	\\
RuO$_2$	&	20.0	&	138.0	&	267.6	&	20.8	&	13.5	&	14.2	&	125.6	&	231.5	&	19.0	&	13.5	&	14.1	\\
NbC	&	18.0	&	201.8	&	298.7	&	30.5	&	25.2	&	25.2	&	157.8	&	266.0	&	23.8	&	18.0	&	18.3	\\
AlN	&	18.0	&	122.2	&	195.1	&	18.5	&	16.2	&	16.2	&	126.7	&	191.2	&	19.1	&	18.0	&	17.8	\\
NbN	&	17.0	&	129.1	&	305.9	&	19.5	&	9.5	&	10.8	&	107.1	&	286.9	&	16.2	&	6.7	&	8.2	\\
HfN	&	17.0	&	158.7	&	269.5	&	24.0	&	17.9	&	18.2	&	132.9	&	248.5	&	20.1	&	13.8	&	14.4	\\
GaN	&	15.1	&	106.0	&	175.3	&	16.0	&	14.0	&	14.1	&	105.5	&	173.7	&	15.9	&	14.0	&	14.1	\\
ZrO$_2$	&	13.0	&	102.0	&	233.5	&	15.4	&	8.4	&	9.5	&	101.8	&	194.8	&	15.4	&	11.0	&	11.6	\\
Si	&	12.0	&	62.5	&	89.1	&	9.4	&	11.8	&	11.5	&	51.2	&	81.3	&	7.7	&	8.6	&	8.8	\\
GaP	&	9.5	&	51.3	&	78.8	&	7.7	&	9.1	&	9.2	&	37.5	&	74.0	&	5.7	&	4.5	&	5.5	\\
AlP	&	9.4	&	46.4	&	82.9	&	7.0	&	6.6	&	7.2	&	45.7	&	82.6	&	6.9	&	6.4	&	7.0	\\
InN	&	9.0	&	54.9	&	124.4	&	8.3	&	5.0	&	6.2	&	50.0	&	115.5	&	7.5	&	4.4	&	5.7	\\
Ge	&	8.8	&	46.2	&	61.5	&	7.0	&	10.5	&	10.0	&	45.5	&	60.3	&	6.9	&	10.4	&	10.0	\\
GaAs	&	7.5	&	40.8	&	62.7	&	6.2	&	7.6	&	7.8	&	34.0	&	58.3	&	5.1	&	5.4	&	6.1	\\
Y$_2$O$_3$	&	7.5	&	62.3	&	137.9	&	9.4	&	5.9	&	6.9	&	60.7	&	123.8	&	9.2	&	6.6	&	7.5	\\
InP	&	5.4	&	31.6	&	60.4	&	4.8	&	4.1	&	5.1	&	25.7	&	58.5	&	3.9	&	2.1	&	3.6	\\
AlAs	&	5.0	&	39.3	&	67.4	&	5.9	&	6.1	&	6.7	&	36.3	&	64.6	&	5.5	&	5.3	&	6.1	\\
GaSb	&	4.5	&	29.6	&	47.0	&	4.5	&	5.4	&	6.0	&	27.2	&	44.4	&	4.1	&	4.8	&	5.5	\\
AlSb	&	4.0	&	28.5	&	49.4	&	4.3	&	4.5	&	5.3	&	26.5	&	47.9	&	4.0	&	3.8	&	4.8	\\
InAs	&	3.8	&	26.2	&	63.6	&	4.0	&	1.8	&	3.4	&	26.1	&	59.4	&	3.9	&	2.1	&	3.6	\\
InSb	&	2.2	&	20.1	&	38.1	&	3.0	&	2.5	&	3.7	&	17.4	&	36.5	&	2.6	&	1.4	&	3.0	\\
ZnS	&	1.8	&	33.9	&	71.2	&	5.1	&	3.6	&	4.8	&	30.9	&	68.8	&	4.7	&	2.8	&	4.2	\\
ZnSe	&	1.4	&	27.5	&	58.2	&	4.2	&	2.8	&	4.1	&	24.9	&	56.7	&	3.8	&	2.0	&	3.5	\\
ZnTe	&	1.0	&	22.1	&	43.8	&	3.3	&	2.5	&	3.8	&	18.4	&	37.3	&	2.8	&	1.8	&	3.2	\\
\hline
\end{tabular}
\end{table*}

\begin{figure*}
    \centering
    \includegraphics[width=1.5\columnwidth]{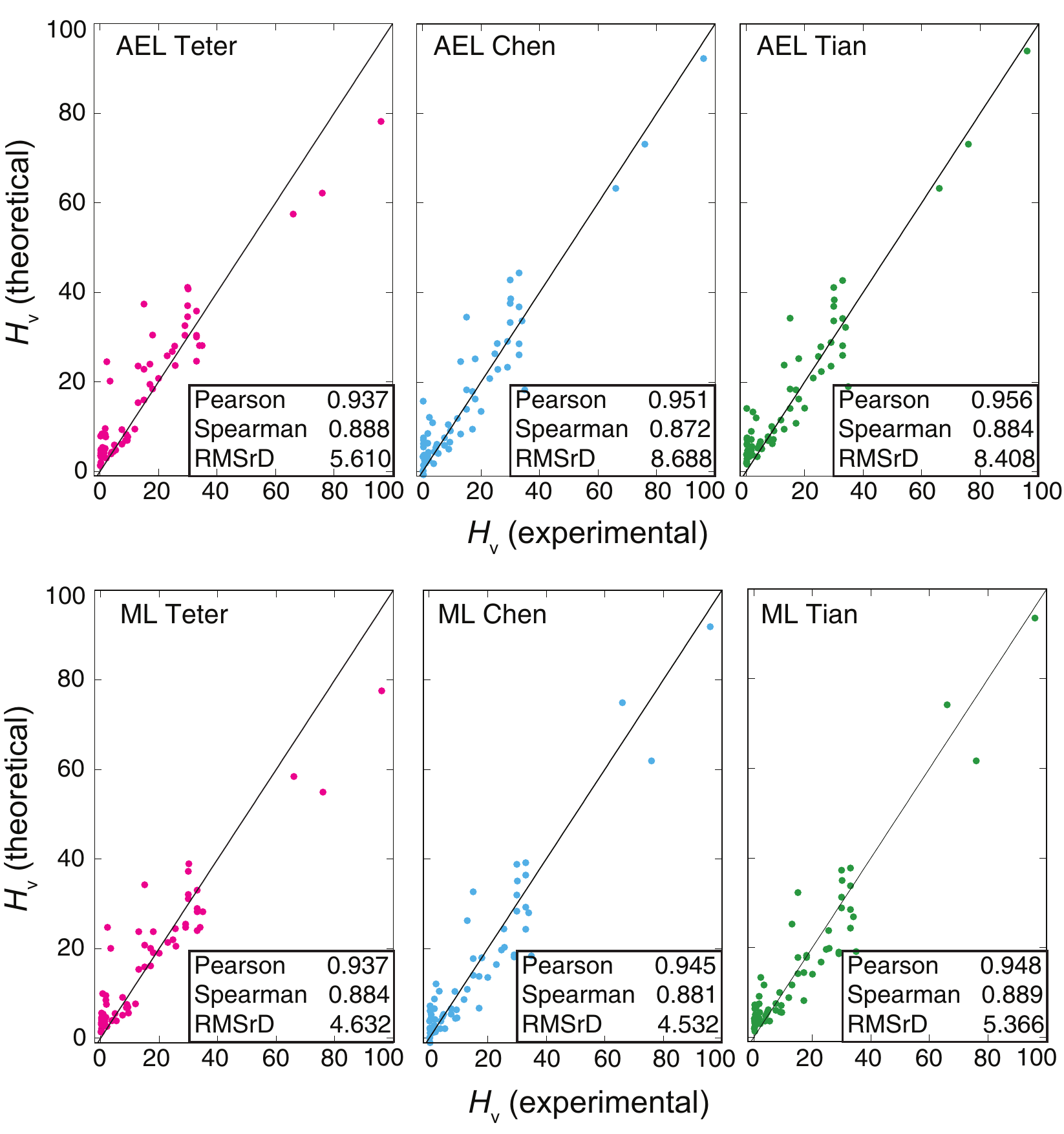}
    \caption{\label{fig:ML_AEL_EXP} A comparison of the Vickers hardness, $H_\text{v}$ in GPa, for 64 materials (see the SI) computed via the Teter (Eq.\ \ref{eqn:TeterModelVickersHardness}), Chen (Eq.\ \ref{eqn:ChenModelVickersHardness}), and Tian (Eq.\ \ref{eqn:TianModelVickersHardness}) models using the Voigt-Reuss-Hill averages of the bulk and shear moduli obtained from the Automatic Elasticity Library (AEL) and via a Machine Learning (ML) model trained on the AFLOW database, with experiment. The line represents a perfect correlation. The Pearson and Spearman coefficients between the experimental and theoretical data are provided, as is the root-mean-squared relative deviation, RMSrD.}
\end{figure*}

To analyze the agreement between the Vickers hardnesses computed via AFLOW-AEL and the three different models in more detail, a correlation analysis was performed, and the results are provided in Fig.\ \ref{fig:ML_AEL_EXP}, which also plots the theoretical vs.\ the experimental data. The Pearson and Spearman coefficients measure the linear correlation between two variables, and the monotonicity of the relation between them, respectively. 
The normalized root-mean-squared relative deviation, RMSrD, between the experimental and computed values was also found, as it can be useful for distinguishing between different methods that may have similar correlations with, but different deviations from, experiment. Whereas the Pearson and Spearman coefficients should ideally be equal to 1, a lower RMSrD is indicative of a better agreement between theory and experiment. All three models showed a high correlation, though the Pearson coefficient was somewhat closer to unity for the Chen and Tian models, as compared to the Teter model. The RMSrD is clearly lower for the Teter model than the other two for this data set when AEL moduli are used. However, the Teter model tends to underestimate the Vickers hardnesses of materials with $H_\text{v}>40$~GPa. Many of the systems in the full data set are soft, with $H_\text{v} < $~5~GPa, and, as previously mentioned, the macroscopic models are not so good at predicting their hardness values. A correlation analysis was also carried out for the subset of materials studied by Chen \cite{Chen:2011a}, see Table \ref{tab:Compare_hard}, and the results are provided in the SI. The correlation coefficients for this subset of systems are higher than for the full data set, and the RMSrD values are significantly lower, $<1$. 

The overwhelming majority of structures generated in a CSP search will not be found within a repository such as AFLOW. Moreover, it would not be feasible to carry out the first-principles geometry optimizations (24 per structure, unless reducible by symmetry) that AFLOW-AEL requires to obtain the bulk and shear moduli for the hundreds, if not thousands, of structures generated within a single CSP search. Therefore, we wondered if it would be possible to take advantage of the power of data science by using bulk and shear moduli obtained from the ML model that was trained on the materials within the AFLOW repository instead \cite{Isayev:2017a}? This Property-Labeled Materials-Fragments model \cite{Isayev:2017a} uses atomic distance and a Voronoi tessellation to characterize the connectivity of a crystal structure: Voronoi cells that are centered on each atom are constructed and two atoms are considered connected if they are within a bonding distance threshold and their Voronoi cells share a face. The connected atoms form a graph, and sections of the graph define the ``materials fragments''. The nodes of the graph are decorated with elemental properties of the atomic species at that site, such as electronegativity, ionization energy, and atomic number. Combinations of these elemental properties, along with the properties of the crystal such as lattice parameters and space group, form the feature vector representing the material. The ML model is based on gradient boosted decision trees \cite{Friedman:2001a}, and is trained on elastic properties calculated using AEL. This model was able to predict the $G_{\sVRH}$ and $B_{\sVRH}$ of over 85\% of the systems tested to within 20~GPa of the values calculated via AEL, with the ML moduli typically being smaller. The deviation was most pronounced for systems with a large $G_{\sVRH}$, and significant differences for $B_{\sVRH}$ were found for graphite and  two theoretical high-pressure B-N phases.  

While it would be preferable to train the ML model directly to predict hardness values, this would require training directly on experimental data, since hardness cannot be calculated directly from first-principles \cite{Tian:2012a,Gao:2010a,Chen:2011a}. However, because experimental hardness data is not readily available in the quantity and quality required to train ML models (reported experimental values can vary by more than 10\% \cite{Gao:2010a}), models based on calculated elastic moduli must be used. For example, a recent combined theoretical/experimental ML directed search for superhard materials trained an ML model to predict $G$ and $B$, which were assumed to be good proxies for hardness, and no attempts to estimate hardness were made \cite{Tehrani:2018a}.

Unsurprisingly, Table \ref{tab:Compare_hard} shows that the AFLOW-ML model also typically underestimated the bulk and shear moduli of the structures considered herein, and it tended to do a better job in predicting $B_{\sVRH}$ as opposed to $G_{\sVRH}$. This suggests that the ML-derived Vickers hardnesses will be smaller than those obtained using AEL for all three models. Comparison of the hardness values obtained with ML vs.\ AEL in Table \ref{tab:Compare_hard} and Fig.\ \ref{fig:ML_AEL_EXP} shows that in general these expectations hold, with the exception of c-BN and AlN for all three models, and SiO$_2$, ZrO$_2$, and Y$_2$O$_3$ for the Chen and Tian models. The correlation coefficients for the ML and AEL results shown in Fig.\ \ref{fig:ML_AEL_EXP} are similar, but the RMSrD values are somewhat smaller for the results obtained using the ML moduli.

Our findings suggest that the ML moduli can be used to obtain reasonable estimates of the hardness of a given material quickly, making it possible to employ them to determine the fitness of an individual in an EA based structure search. But which one of the macroscopic models should be used? During our EA searches on the elemental carbon system we found that in some instances the ML-derived $B_{\sVRH}$ were extremely small, and when they were used in the denominator in Eqs.\ \ref{eqn:ChenModelVickersHardness} and \ref{eqn:TianModelVickersHardness}, unrealistically hard values were obtained, with several systems being predicted to be much harder than diamond itself. Only the Teter model was able to predict that these crystals had a very low Vickers hardness. For the 64 systems in our test set, the Pearson and Spearman correlation coefficients between the Vickers hardness values calculated using the Teter model and AEL vs.\ ML were 0.993 and 0.980, respectively, and the RMSrD was 0.156. Moreover, the Chen and Tian models are theoretically less satisfying than the Teter model, since they are empirically fitted and result in an incorrect dimension. Thus, the Teter model, Eq.\ \ref{eqn:TeterModelVickersHardness}, was chosen for the hardness evaluations since it can differentiate between the hard crystals that are kept in the gene pool during the evolutionary search, as described in the following section, and those that are soft.

\subsection{The \textsc{XtalOpt} Evolutionary Algorithm}
EAs employ concepts from biological evolution to find an optimal solution for problems that have many degrees of freedom. When applied towards \emph{a priori} CSP, EAs search for the lattice parameters and atomic coordinates that minimize or maximize a computed quantity. Because EAs are typically concerned with finding the global minimum (along with important local minima)  on the PES, they attempt to minimize the computed energy/enthalpy \cite{Zurek:2014d}. An EA starts by generating a set of random structures (structures chosen by the user, also known as seeds, may also be employed) that are relaxed to the nearest local minima by an external program. The fitness of each optimized individual is calculated and used to determine the probability that it will be chosen for procreation. Child structures are created either via a two-parent breeding operation, or by a mutation of a single parent.  Further details about the \textsc{XtalOpt} EA, and its subprograms can be found in Refs.\ \citen{Lonie:2011a,Zurek:2017k,Zurek:2018j,xtalopt,Zurek:2011i,Zurek:2016h}. 

The original implementation of the open source EA for CSP, \textsc{XtalOpt} \cite{Lonie:2011a}, uses roulette wheel selection where the probability $p_i$ that a structure with energy/enthalpy $E_i$ is chosen for procreation is equal to its fitness. The probability is calculated as: 
\begin{equation}\label{eqn:XtalOptEnthalpyFitnessFunction}
p_i = N\left(1 - \frac{E_i - E_\text{min}}{E_\text{max} - E_\text{min}}\right),
\end{equation}
where $E_\text{min}$ and $E_\text{max}$ are the lowest and highest energies/enthalpies in the breeding pool, respectively, and $N$ is a normalization constant chosen to ensure that $\sum p_i = 1$. The lower the energy/enthalpy of an individual, the more likely it is to be selected as a parent for the subsequent generation.

In the algorithms designed to predict superhard materials that are implemented within CALYPSO \cite{Zhang:2013a} and USPEX \cite{Lyakhov:2011a}, the hardness, as calculated via a microscopic model, is employed to determine an individual's fitness. Using roulette wheel selection, and maximizing the Vickers hardness instead of minimizing the energy/enthalpy, Eq.\ \ref{eqn:XtalOptEnthalpyFitnessFunction} becomes
\begin{equation}\label{eqn:XtalOptHardnessFitnessFunction}
p_i = N\left(1 - \frac{H_\text{v,max} - H_{\text{v},i}}{H_\text{v,max} - H_\text{v,min}}\right).
\end{equation}
The higher the hardness of an individual, $H_{\text{v},i}$, the more likely it is to be selected for breeding.

Because we are interested in predicting the structures of superhard materials that could potentially be synthesized, it is desirable that they correspond to low lying local minima. Therefore, the new fitness function implemented within \textsc{XtalOpt} combines Eqs.\ (\ref{eqn:XtalOptEnthalpyFitnessFunction}) and (\ref{eqn:XtalOptHardnessFitnessFunction}) to favor the selection of structures that are both low in enthalpy and high in hardness via:
\begin{multline}\label{eqn:XtalOptCombinedFitnessFunction}
p_i = N\Bigg[1 - w\left(\frac{H_\text{v,max} - H_{\text{v},i}}{H_\text{v,max} - H_\text{v,min}}\right) - \\ (1 - w)\left(\frac{E_i - E_\text{min}}{E_\text{max} - E_\text{min}}\right)\Bigg],
\end{multline}
where the weight, $w$, is a fractional number between 0 and 1. 
Herein, we employed a $w=0.5$, and the Vickers hardnesses used to determine $p_i$ were computed via the Teter model, Eq.\ \ref{eqn:TeterModelVickersHardness}, using the ML values of the shear modulus. 

Most EAs do not keep every optimized structure within the breeding pool, but rather only a user-specified number of the fittest individuals. In the original version of \textsc{XtalOpt}, the lowest energy/enthalpy structures were placed in the breeding pool. In the current implementation, the fitness of each individual is computed prior to normalization using Eq.\ \ref{eqn:XtalOptCombinedFitnessFunction}. The crystals with the highest $p_i$ are placed in the
breeding pool, and their normalized probabilities are computed and used.

\subsection{Applications to Carbon\label{sec:smallcarbon}}
Above we have shown that the Vickers hardness of a compound can be computed with reasonable accuracy via Teter's equation using the shear modulus obtained with a ML algorithm, and proposed a fitness function that can be used in conjunction within an evolutionary structure search to predict superhard materials. To determine how well this approach works, we performed an EA search on elemental carbon. This system was chosen because of the diversity of the crystalline carbon family that arises from the possibility of $sp$, $sp^2$ and $sp^3$ hybridized bonds, leading to an infinite number of structures within its chemical space. Moreover, the hardness of its family members ranges from that of  soft layered graphite to diamond, the hardest known material on this planet, as well as many well-known superhard species including lonsdaleite, M-carbon, Z-carbon, F-carbon, among many others.
The wide range of possible Vickers hardnesses and crystalline topologies within the carbon system creates a tremendous space for \textsc{XtalOpt} to explore, and we are able to evaluate its performance by observing its efficiency on finding pearls, the superhard species, from the ocean of carbon crystals. We expect our EA search will find many reported structures, illustrating that it works, and show that it is a predictive tool that can discover new topologies.

\begin{figure}[!ht]
    \centering
    \includegraphics[width=\columnwidth]{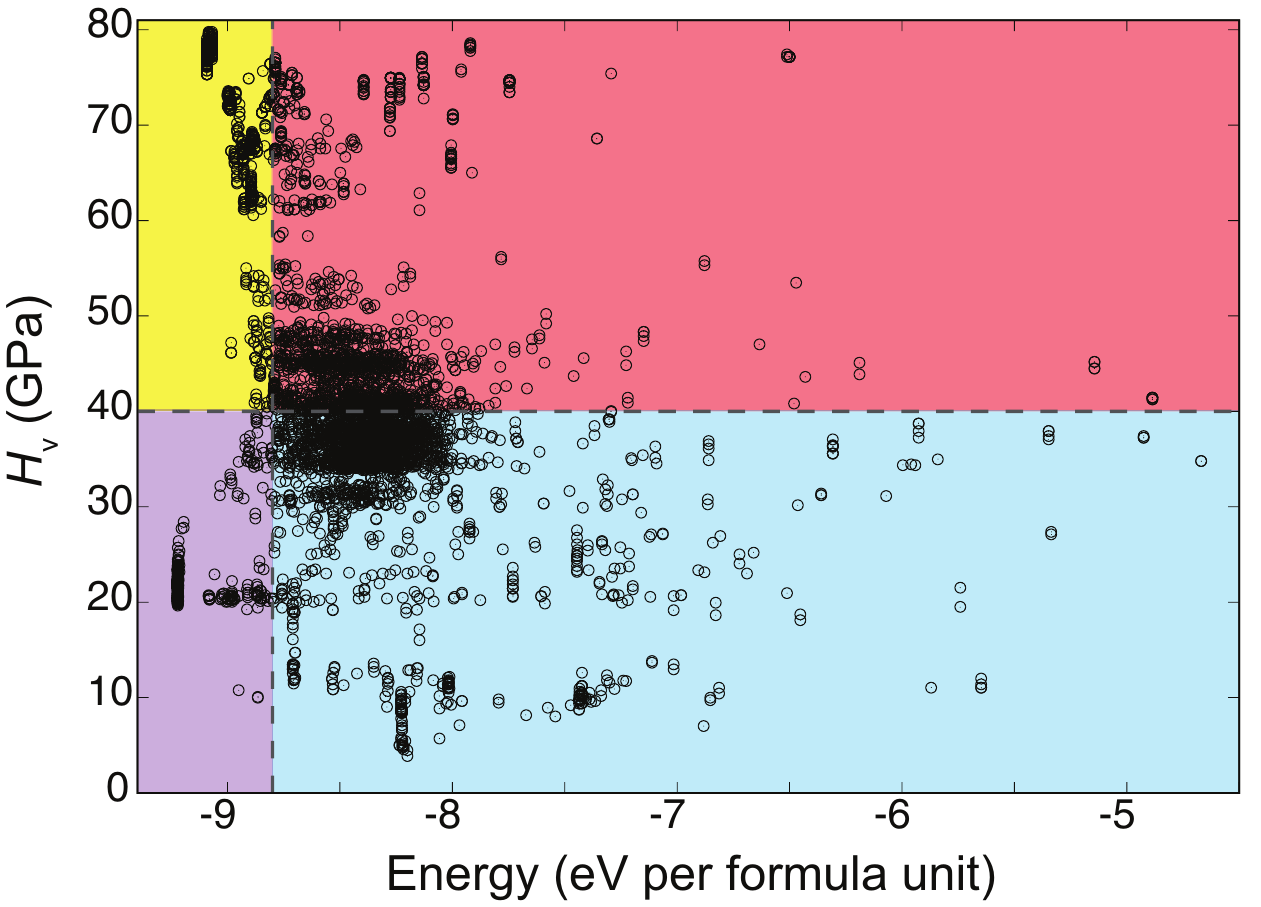}
    \caption{\label{fig:eh_map}The hardness vs.\ energy map of all carbon topologies predicted by the \textsc{XtalOpt} searches. The horizontal line corresponds to a $H_\text{v,Teter}^{\text{ML}}=40$~GPa, and the vertical line to an energy of $-8.80$~eV/atom. The structures in the yellow quadrant are both stable and superhard.}
    \label{fig:ga}
\end{figure}

Fig.\ \ref{fig:eh_map} plots $H_\text{v,Teter}^{\text{ML}}$ for all of the 5624 optimized individuals vs.\ their energy. The plot is partitioned into four quadrants following the standard suggested by Zhang \textit{et al.} \cite{Zhang:2013a}. We define superhard structures to have a Vickers hardness larger than 40~GPa, and thermodynamically stable structures to be those whose 0~K energies are less than $-8.80$~eV/atom, which is within $0.30$~eV/atom of diamond. For comparison, the energy of M-carbon, which has been synthesized under pressure \cite{mao2003bonding,Wang:2012a,Li:2009a}, was computed to be $-8.93$~eV/atom. The superhard species that could potentially be synthesized lie in the top left quadrant of the hardness-energy map, and soft stable ones are found in the bottom left. Superhard phases with comparably high energies are located in the top right hand side of the plot, and unstable soft systems are at the bottom right. The energy of 1324 structures was less than $-8.80$~eV/atom, and from these 827 had a Vickers hardness greater than 40~GPa. Duplicates were removed from the yellow quadrant in Fig.\ \ref{fig:ga}, resulting in 89 distinct topologies, which were found by \textsc{XtalOpt}, that are both hard and thermodynamically favored. Phonon calculations showed that ten of these phases, all containing $sp^2$ carbons, were dynamically unstable. Therefore, in our searches, which employed cells containing 8, 12, 16 and 20 carbon atoms, 79 hard and stable individuals were found.

To determine if the predicted structures are already known, and avoid making false claims of novelty, the Samara Carbon Allotrope Database (SACADA) was employed \cite{sacada-web,hoffmann2016homo}. SACADA has collected the crystal structures and physical properties from both experiment and theory of more than 500 3D carbon allotropes. Among the hard species found via \textsc{XtalOpt}, 36 are already known, including diamond (\#1, \textbf{dia}), lonsdaleite (\#37, \textbf{lon}), a chiral-framework (\#29, \textbf{unj}), M-carbon (\#224, \textbf{cbn}), Z-carbon (\#141, \textbf{sie}), F-carbon (\#225, 4$^4$T35), bct4-carbon (\#60, \textbf{crb}), and BC$_8$-carbon (\#20, \textbf{gsi}). Some of these are also present in the AFLOW crystal prototype library \cite{mehl:2017a}. The SACADA ID numbers and densities of the previously known phases are reported in Table \ref{tab:carbon_reported}, which also provides a comparison between our $H_\text{v,Teter}^{\text{ML}}$ results with the hardness values computed by others.  The reason why the $H_\text{v,Teter}^{\text{ML}}$ for diamond given in Table \ref{tab:Compare_hard} differs slightly from that in Table \ref{tab:carbon_reported} is because in the first case the structure was taken as-is from the AFLOW database, and in the second case it was optimized again by us. 

Fig.\ \ref{fig:ML_AEL_EXP} clearly illustrates that the Teter model tends to underestimate the Vickers hardnesses of superhard materials. Therefore, it is not surprising that almost all of our $H_\text{v,Teter}^{\text{ML}}$ values are lower than the Vickers hardness estimates of others. However, once a structure has been predicted by an EA, it is always possible to compute its hardness using a more accurate and computationally expensive hardness model. What is important is that for the unit cell sizes employed herein, our EA is able to find most of the hard stable carbon structures predicted earlier in Refs.\ \citen{Zhang:2013a} and \citen{Lyakhov:2011a}. 
Plus, we find many new stable superhard phases not reported in SACADA, or known as of yet in the literature.

\begin{table*}[!ht]
    \setlength{\tabcolsep}{9pt}
    \renewcommand{\arraystretch}{1.2}
    \centering
\caption{Stable superhard structures discovered using the \textsc{XtalOpt} evolutionary algorithm that have been predicted before. Either the well-known names, or those used in the first paper reporting the structure, are provided, along with the SACADA ID and topology. The Vickers hardness (GPa) as calculated via the Teter model using the ML shear modulus, $H_\text{v,Teter}^{\text{ML}}$, or computed in other studies, $H_\text{v}^{\text{others}}$, as well as the density, $\rho$, in g/cm$^3$ of the optimized structures is also given. $^\dagger$Structures are in the SACADA database but have not yet been published online.} \label{tab:carbon_reported}
\begin{tabular}{llccc}
\hline\hline
Names                                         	&	SACADA ID (Topology)	&	$H_\text{v,Teter}^{\text{ML}}$ 	&	$H_\text{v}^{\text{others}}$ 	&	 $\rho$ 	\\
\hline
diamond                    	&	1 (\textbf{dia})	&	75.6	& 89.5\footnotemark[1], 87.3\footnotemark[1], 95.4\footnotemark[1], 97.5\footnotemark[1], 89.7\footnotemark[2]	&	3.50	\\
bc8                                             	&	20 (\textbf{gsi})	&	74.4	&	88.6\footnotemark[1]	&	3.56	\\
chiral-framework                             	&	29 (\textbf{unj})	&	47.1	&	81.5\footnotemark[1], 90.8\footnotemark[1], 81.3\footnotemark[2]	&	3.21	\\
Lonsdaleite                                     	&	37 (\textbf{lon})	&	79.3	&	 89.1\footnotemark[1], 89.1\footnotemark[2], 97.3\footnotemark[3]	&	3.49	\\
bct-C$_4$                                     	&	60 (\textbf{crb})	&	65.6	&	84.4\footnotemark[1], 82.0\footnotemark[1], 92.9\footnotemark[1], 84.0\footnotemark[2]	&	3.31	\\
Y-carbon                        	&	81 (\textbf{cag})	&	62.8	&		&	3.31	\\
T12                                             	&	107 (\textbf{cdp})	&	68.2	&	85.1\footnotemark[1]	&	3.35	\\
4H-diamond                                     	&	111 (\textbf{cfc})	&	78.3	&	98.1\footnotemark[3]	&	3.50	\\
bikitaite carbon               	&	124 (\textbf{bik})	&	48.4	&		&	3.23	\\
Z-carbon                                     	&	141 (\textbf{sie})	&	71.8	&	85.1\footnotemark[1], 84.4\footnotemark[1], 95.1\footnotemark[1]	&	3.40	\\
6B                              	&	179 (NSI)	&	48.8	&		&	3.25	\\
$C2/m$-16                       	&	186 (4$^3$T85)	&	68.6	&	84.9\footnotemark[1]	&	3.35	\\
12R-diamond                  	&	209 (SiC12)	&	77.0	&		&	3.50	\\
oc16                          	&	210 (3,4$^3$T72)	&	52.3	&	42.9\footnotemark[3]	&	3.27	\\
W-carbon                                     	&	218 (\textbf{cnw})	&	70.4	&	85.3\footnotemark[1], 83.1\footnotemark[1], 93.8\footnotemark[1]	&	3.35	\\
M-carbon                                     	&	224 (\textbf{cbn})	&	68.0	&	85.0\footnotemark[1], 82.7\footnotemark[1], 93.5\footnotemark[1], 84.3\footnotemark[2]	&	3.34	\\
F-carbon                                     	&	225 (4$^4$T35)	&	70.4	&	84.7\footnotemark[1]	&	3.33	\\
P2$_1$/m                         	&	245 (4$^5$T27)	&	70.8	&	85.4\footnotemark[1]	&	3.36	\\
12D                             	&	256 (4$^6$T15)	&	72.1	&		&	3.40	\\
$oP20$                            	&	260 (4$^6$T16)	&	75.4	&	91.4\footnotemark[3]	&	3.42	\\
Z-carbon-3                    	&	262 (4$^6$T7)	&	72.1	&	92.6\footnotemark[3]	&	3.40	\\
$mS32$                           	&	291 (4$^8$T16)	&	72.6	&	90.8\footnotemark[3]	&	3.41	\\
Z-carbon-2\                     	&	293 (4$^8$T5)	&	75.6	&	92.4\footnotemark[3]	&	3.42	\\
$mP16$                           	&	294 (4$^8$T15)	&	75.8	&	91.0\footnotemark[3]	&	3.41	\\
16D                             	&	336 (4$^{16}$T3)	&	75.9	&		&	3.41	\\
16B                             	&	337 (4$^{16}$T2)	&	72.7	&		&	3.42	\\
G6\                             	&	358 (4$^6$T28)	&	66.8	&		&	3.33	\\
G21                            	&	365 (4$^5$T49)	&	69.3	&		&	3.33	\\
G120                           	&	420 (4$^4$T85)	&	68.7	&		&	3.36	\\
G153                           	&	447 (\textbf{cfe})	&	77.8	&		&	3.50	\\
G158                           	&	451 (mbc-4,4-$Imma$)	&	47.1	&		&	3.26	\\
G178                           	&	470 (4$^8$T44)	&	78.3	&		&	3.48	\\
G225                          	&	511 (4$^8$T49)	&	72.6	&		&	3.41	\\
3,4,4T154$^\dagger$                     	&	 not published       	&	48.0	&		&	3.19	\\
8-layered SiC polytype$^\dagger$         	&	 not published     	&	77.0	&		&	3.50	\\
Deem hyp. Zeolite             	&	 8047698     	&	54.5	&		&	3.30	\\
\hline
\end{tabular}
    \footnotetext[1]{Hardness values are from Ref.\ \citen{Zhang:2013a} and references therein.}
    \footnotetext[2]{Hardness values are from Ref.\ \citen{Lyakhov:2011a}.}
    \footnotetext[3]{Hardness values are from the SACADA database.}
\end{table*}

Table \ref{tab:carbon_reported} shows that the computed $H_\text{v,Teter}^{\text{ML}}$ for the previously known individuals is typically 10-20~GPa lower than the values obtained using the microhardness models implemented within CALYPSO \cite{Zhang:2013a} and USPEX \cite{Lyakhov:2011a}. One exception is for the chiral-framework structure (\#29, \textbf{unj}) originally predicted in Ref.\ \citen{Pickard:2010a}, which we find to be 28.5~GPa less hard than diamond. The aforementioned microhardness models computed diamond to be only 8-8.4~GPa harder than this $P6_522$ symmetry structure. It should be noted, however, that the chiral framework is significantly less dense than diamond ($\rho=$~3.21~g/cm$^3$ vs.\ 3.50~g/cm$^3$), and it would be quite remarkable if it were nearly as hard. The value of $H_\text{v,Teter}^{\text{ML}}$ obtained herein for this phase, 47.1~GPa, is close to Vickers hardnesses we calculate for phases with a similar density: bikitaite carbon, 6B, G158 and 3,4,4T154.

In addition to the previously predicted phases, our algorithm identified 53 new low-energy topologies that were superhard. Calculations of the phonons and elastic constants confirmed that 43 of these were dynamically and mechanically stable. A summary table (summary.xlsx), which lists the hardness, relative stability and topological properties of these phases is provided in the SI, and their fascinating structures, properties and electronic structures will be discussed in detail in a follow-up paper. Table \ref{tab:new_topologies} lists the ML and AEL values of their bulk and shear moduli, as well as the corresponding Vickers hardnesses computed with the Teter model. Their energy relative to diamond, density, and pressure above which their enthalpies become lower than that of graphite are also given. The caption of Table \ref{tab:new_topologies} provides the correlation coefficients and RMSrD between the ML and AEL Vickers hardnesses computed for these new phases using the Teter, Chen and Tian equations. The best agreement between the ML and AEL values is obtained for the Teter model.

\begin{table*}
\setlength{\tabcolsep}{9pt}
\renewcommand{\arraystretch}{1.2}
\caption{New stable superhard structures predicted by the \textsc{XtalOpt} 
evolutionary algorithm. 
$E_{\text{ref}}$ is the energy relative to diamond (eV/atom). The Voigt-Reuss 
Hill (VRH) average of the bulk, $B$, and shear, $G$, moduli obtained from the 
Automatic Elastic Library (AEL) and via a Machine Learning (ML) model trained on 
the AFLOW database, and the Vickers hardness, $H_\text{v}$, as computed via the 
Teter model (Eq.\ \ref{eqn:TeterModelVickersHardness}) are provided. The 
density, $\rho$, in g/cm$^3$, percent of diamond and lonsdaleite found in the structure, and pressure, $P$, above which the structures are 
computed to become more stable than graphite are listed.  $B$, $G$, $H_\text{v}$, and 
$P$ are given in units of GPa. The Pearson and Spearman coefficients, along with 
the RMSrD between the AEL and ML results are 0.858, 0.895, 0.115 for the Teter 
model, 0.811, 0.849, 0.201 for the Chen model and 0.818, 0.854 and 0.200 for the 
Tian model.}
\label{tab:new_topologies}
\begin{center}
\begin{tabular}{lcccccccccc}
\hline\hline
Name	&	$E_{\text{ref}}$	&	$B^{\text{ML}}_{\sVRH}$	&	
$B^{\text{AEL}}_{\sVRH}$	&	$G^{\text{ML}}_{\sVRH}$	&	
$G^{\text{AEL}}_{\sVRH}$	&	$H_\text{v,Teter}^{\text{ML}}$	&	
$H_\text{v,Teter}^{\text{AEL}}$	&	$\rho$	&	\%\textbf{dia}/\textbf{lon}	
&	$P$	\\
\hline
$R\overbar{3}m$-16	&	0.001	&	404.5	&	434.7	&	
504.9	&	521.6	&	76.2	&	78.8	&	3.50	&	
100	&	7.6	\\
$R3m$-16	&	0.007	&	404.6	&	435.1	&	504.8	
&	523.0	&	76.2	&	79.0	&	3.50	&	100	
&	8.1	\\
$P2/m$-12	&	0.133	&	421.6	&	414.6	&	494.2	
&	467.1	&	74.6	&	70.5	&	3.38	&	31.6	
&	20.3	\\
$Imm2$-12	&	0.137	&	387.0	&	415.2	&	474.3	
&	468.1	&	71.6	&	70.7	&	3.38	&	31.9	
&	20.7	\\
$C2/m$-20c	&	0.142	&	392.1	&	419.7	&	468.4	
&	487.7	&	70.7	&	73.6	&	3.40	&	38.3	
&	20.6	\\
$Imma$-16	&	0.143	&	387.2	&	413.4	&	468.8	
&	452.8	&	70.8	&	68.4	&	3.38	&	72.9	
&	21.2	\\
$P\overbar{1}$-16a	&	0.149	&	399.0	&	413.3	&	
447.9	&	461.3	&	67.6	&	69.7	&	3.36	&	
0	&	22.4	\\
$C2/m$-20d	&	0.160	&	380.5	&	406.4	&	424.3	
&	453.0	&	64.1	&	68.4	&	3.32	&	38.0	
&	24.4	\\
$Pmm2$-20	&	0.165	&	401.1	&	406.9	&	424.0	
&	454.3	&	64.0	&	68.6	&	3.32	&	38.3	
&	25.0	\\
$C2/m$-20a	&	0.176	&	386.5	&	409.1	&	447.7	
&	440.2	&	67.6	&	66.5	&	3.35	&	19.0	
&	25.3	\\
$C2/m$-16a	&	0.179	&	378.5	&	406.1	&	439.1	
&	419.8	&	66.3	&	63.4	&	3.33	&	11.9	
&	26.1	\\
$C2/m$-16b$^\dagger$	&	0.187	&	382.1	&	409.1	&	
449.1	&	403.9	&	67.8	&	61.0	&	3.35	&	
24.3	&	26.3	\\
$C2/m$-20b	&	0.193	&	384.7	&	410.5	&	450.0	
&	445.2	&	67.9	&	67.2	&	3.36	&	20.1	
&	26.7	\\
$C2/m$-12a	&	0.193	&	384.9	&	411.0	&	455.3	
&	448.3	&	68.7	&	67.7	&	3.36	&	0	
&	26.5	\\
$C2/m$-16c	&	0.207	&	382.8	&	411.4	&	474.3	
&	448.4	&	71.6	&	67.7	&	3.38	&	48.0	
&	27.6	\\
$C2/m$-12b	&	0.211	&	370.4	&	396.1	&	317.0	
&	424.3	&	47.9	&	64.1	&	3.24	&	0	
&	33.1	\\
$P\overbar{1}$-20	&	0.215	&	392.5	&	388.7	&	
303.5	&	403.2	&	45.8	&	60.9	&	3.18	&	
0	&	37.1	\\
$Cmmm$-12a	&	0.216	&	367.7	&	389.6	&	300.0	
&	397.3	&	45.3	&	60.0	&	3.18	&	0	
&	36.9	\\
$C2/m$-12c	&	0.226	&	370.4	&	391.7	&	314.3	
&	392.6	&	47.5	&	59.3	&	3.23	&	0	
&	35.9	\\
$Cmmm$-20$^\dagger$	&	0.227	&	394.4	&	400.4	&	
346.8	&	396.1	&	52.4	&	59.8	&	3.28	&	
19.3	&	33.2	\\
$C2/m$-16d	&	0.228	&	377.4	&	401.9	&	363.8	
&	414.7	&	54.9	&	62.6	&	3.30	&	11.4	
&	32.7	\\
$P2_1/m$-16a	&	0.228	&	414.2	&	404.5	&	440.0	
&	444.6	&	66.4	&	67.1	&	3.32	&	12.7	
&	32.1	\\
$Cmmm$-12b$^\dagger$	&	0.233	&	381.8	&	386.1	&	
295.1	&	319.8	&	44.6	&	48.3	&	3.17	&	
0	&	39.7	\\
$P\overbar{1}$-16c$^\dagger$	&	0.242	&	407.2	&	412.6	
&	472.1	&	445.0	&	71.3	&	67.2	&	3.38	
&	48.4	&	31.2	\\
$Pmma$-16a	&	0.242	&	394.3	&	399.0	&	360.6	
&	408.8	&	54.5	&	61.7	&	3.29	&	23.8	
&	34.6	\\
$Pmma$-12	&	0.242	&	396.4	&	391.3	&	330.1	
&	401.6	&	49.8	&	60.6	&	3.22	&	0	
&	38.8	\\
$P\overbar{1}$-16d	&	0.248	&	411.5	&	421.8	&	
504.5	&	480.3	&	76.2	&	72.5	&	3.45	&	
61.6	&	29.8	\\
$Pm$-16	&	0.250	&	410.2	&	399.2	&	375.8	&	
411.9	&	56.8	&	62.2	&	3.29	&	0	&	
35.5	\\
$P2_1$-16	&	0.255	&	404.0	&	408.9	&	451.5	
&	448.1	&	68.2	&	67.7	&	3.35	&	0	
&	33.8	\\
$P\overbar{1}$-16e$^\dagger$	&	0.258	&	408.5	&	412.5	
&	473.3	&	449.1	&	71.5	&	67.8	&	3.38	
&	49	&	33.0	\\
$P\overbar{1}$-16b$^\dagger$	&	0.259	&	410.3	&	412.6	
&	479.2	&	448.9	&	72.4	&	67.8	&	3.38	
&	48.7	&	33.3	\\
$P2_12_12_1$-16	&	0.263	&	389.2	&	389.6	&	311.4	
&	380.3	&	47.0	&	57.4	&	3.22	&	0	
&	41.7	\\
$P1$-12	&	0.267	&	386.9	&	386.7	&	305.5	&	
366.1	&	46.1	&	55.3	&	3.17	&	0	&	
46.5	\\
$Pmma$-16b	&	0.274	&	397.0	&	393.0	&	343.6	
&	392.6	&	51.9	&	59.3	&	3.26	&	46.7	
&	40.4	\\
$P2/c$-16	&	0.276	&	393.8	&	395.8	&	322.5	
&	415.2	&	48.7	&	62.7	&	3.25	&	0	
&	41.9	\\
$P1$-16d	&	0.276	&	412.3	&	414.4	&	479.6	
&	461.7	&	72.4	&	69.7	&	3.40	&	0	
&	34.4	\\
$P1$-16a$^\dagger$	&	0.278	&	403.8	&	406.0	&	
447.0	&	409.9	&	67.5	&	61.9	&	3.35	&	
43.4	&	36.5	\\
$P\overbar{1}$-12	&	0.281	&	417.0	&	424.5	&	
500.4	&	492.7	&	75.6	&	74.4	&	3.47	&	
33.7	&	32.3	\\
$P1$-16b$^\dagger$	&	0.282	&	385.7	&	388.6	&	
300.0	&	372.4	&	45.3	&	56.2	&	3.21	&	
0	&	45.9	\\
$P1$-16c	&	0.285	&	414.6	&	417.2	&	485.9	
&	467.4	&	73.4	&	70.6	&	3.43	&	23.5	
&	34.3	\\
$P\overbar{1}$-16f	&	0.286	&	410.1	&	411.8	&	
476.6	&	451.9	&	72.0	&	68.2	&	3.40	&	
36.7	&	35.5	\\
$P2/m$-16	&	0.288	&	401.3	&	389.5	&	329.0	
&	386.6	&	49.7	&	58.4	&	3.22	&	0	
&	44.8	\\
$P1$-16e	&	0.290	&	413.1	&	417.4	&	486.5	
&	470.2	&	73.5	&	71.0	&	3.42	&	23.7	
&	34.9	\\
\hline
\end{tabular}
\end{center}
$^\dagger$ Structures that contain $sp^2$ carbon atoms.
\end{table*}

M-carbon (\#224, \textbf{cbn}), whose $H_\text{v,Teter}^{\text{ML}}$ is computed to be 68.0~GPa, can be made by compressing graphite to $\sim$17-19~GPa \cite{mao2003bonding,Wang:2012a}.  Thus, it might be possible to synthesize some of the newly predicted superhard phases under pressure by an appropriate choice of the carbon allotrope starting material and the synthesis route. Metastable carbon phases have been previously synthesized by compressing glassy carbon \cite{Shiell:2016a,Hu:2017a} or onion carbon \cite{Huang:2014a,Tao:2017a}, for instance. To study this further, we computed the pressures above which the newly predicted superhard phases become more stable than graphite. The transition pressures for diamond-like structures (more on this below) were typically between 7\textasciitilde8 GPa, whereas other structures have transition pressures within 23\textasciitilde47~GPa. Generally speaking, the higher density phases were predicted to become enthalpically favored over graphite at lower pressures, and they had larger Vickers hardnesses than the low density phases, as expected.

Since both the ML and AEL shear moduli of lonsdaleite are larger than that of diamond, but the bulk moduli are about the same, lonsdaleite is predicted to have a slightly higher Vickers hardness by all three macroscopic hardness models.  Lonsdaleite has, in fact, been previously predicted to have superior mechanical properties as compared to diamond \cite{Pan:2009a,Qingkun:2011a}, but its experimental existence has been questioned \cite{Nmeth:2014a}, although there are some recent claims of its synthesis both through static \cite{Shiell:2016a} and shock compression \cite{Turneaure:2017a}. 

\begin{figure}
    \centering
    \includegraphics[width=\columnwidth]{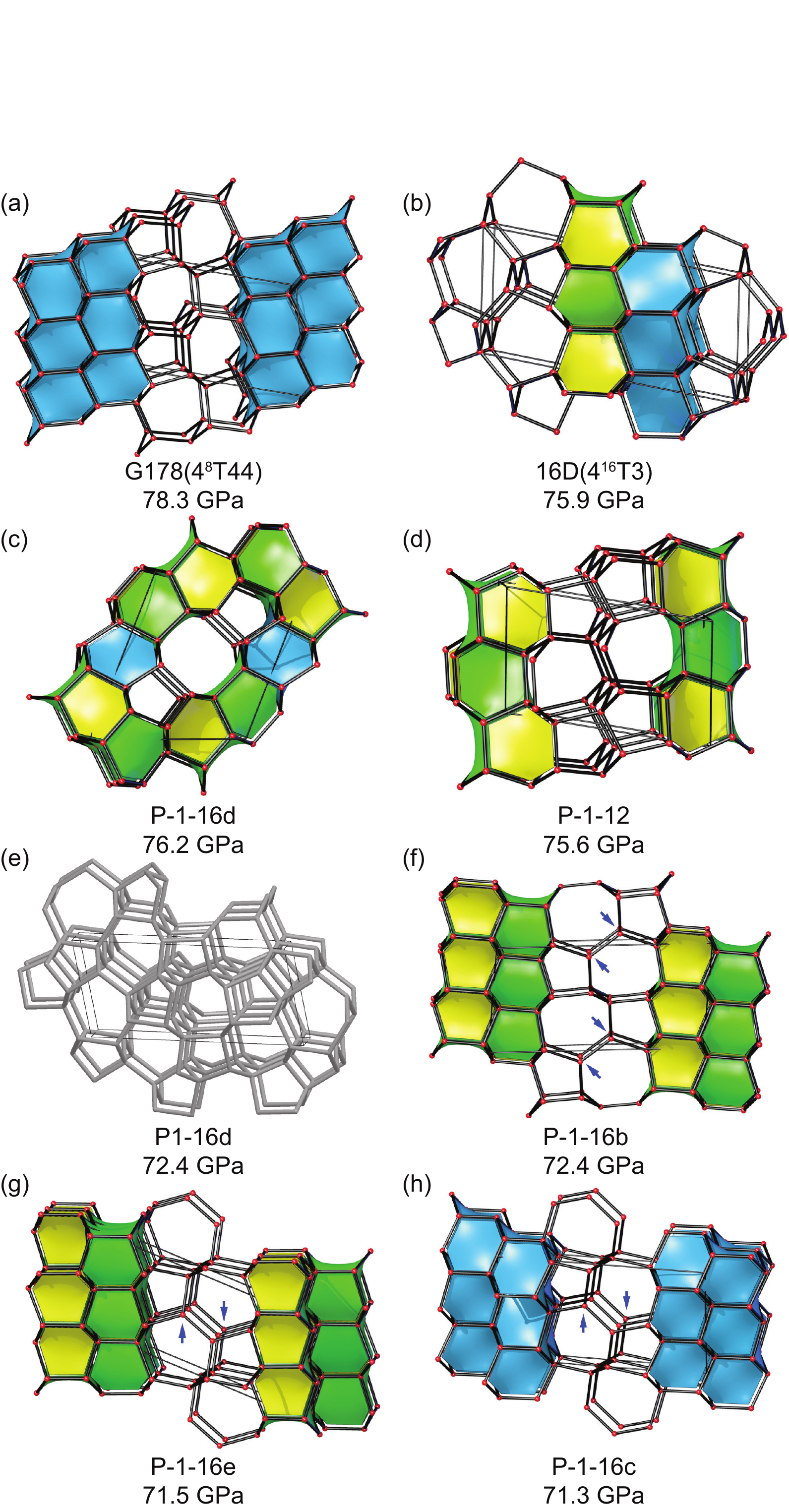}
    \caption{(a,b) Previously known (see Table\ \ref{tab:carbon_reported}), and (c-h) newly predicted (see Table \ref{tab:new_topologies}) superhard phases. Their computed $H_\text{v,Teter}^{\text{ML}}$ are provided. The cages colored in blue are structurally related to diamond (\textbf{dia}), and the cages colored in yellow and green are structurally related to lonsdaleite (\textbf{lon}). The arrows in (f-h) point towards the $sp^2$ carbons. \label{fig:strucs} }
\end{figure}

$H_\text{v,Teter}^{\text{ML}}$ for a few of the predicted structures was computed to be at least as high as that of diamond. This includes the following from Table \ref{tab:carbon_reported}: 4H and 12R diamond, G153, and 8-layered SiC. These are all composed entirely of mixtures of lonsdaleite and diamond: they are members of the infinite family of \textbf{dia}-\textbf{lon} polytypes. The tiling approach, wherein a 3D periodic net is represented as a tiling of generalized polyhedra, has previously been employed to analyze the topology of $sp^3$ bonded and amorphous carbon allotropes \cite{baburin2015zeolite, deringer2017extracting}. Using this method to extract the content of diamond and lonsdaleite topological building blocks in each superhard allotrope found in our EA searches revealed two new \textbf{dia}-\textbf{lon} polytypes that were also, unsurprisingly, found to be harder than diamond ($R\overbar{3}m$-16 and $R3m$-16, see Table \ref{tab:new_topologies}). 

The same analysis technique revealed that G178 (\#470 $4^8T44$), and 16D (\#336 $4^{16}T3$), illustrated in Fig.\ \ref{fig:strucs}(a,b), as well as $mP16$ (\#294 $4^8T16$), shown in Fig.\ 2 in Ref.\ \citen{baburin2015zeolite}, contained more than 47\% of diamond/lonsdaleite tiles (see the supplementary excel table). The newly discovered $P\overbar{1}$-16d and $P\overbar{1}$-12 superhard species that are shown in Fig.\ \ref{fig:strucs}(c,d) contained  61.6\% and 33.7\% diamond/lonsdaleite, respectively, along with tiles found in  $mC12$ (\#147, 4$^2$T112) and $oC16$-I (\#175, 4$^3$T84), respectively. 
The $mC12$ and  $oC16$-I  phases have previously been predicted to be superhard \cite{selli2011superhard}, and herein their $H_\text{v,Teter}^{\text{ML}}$ values were computed to be somewhat lower than that of diamond, 72.6~GPa and 73.5~GPa, respectively.  Thus, the newly predicted species that are estimated to be at least as hard as diamond contain a mixture of tiles found in diamond, lonsdaleite, or other superhard phases. 

The first superhard phase that did not contain diamond or lonsdaleite, $P1$-16d, is shown in Fig.\ \ref{fig:strucs}(e). It is very asymmetric and appears close to amorphous, or disordered, with 16 different atom types. Indeed, it has been shown that disorder in a material can enhance its hardness \cite{sarker:2018a}. The topological analysis further revealed that systems with $sp^2$ hybridized carbon atoms that had an  $H_\text{v,Teter}^{\text{ML}}>$~70~GPa contained a high percentage of diamond and/or lonsdaleite, at least 40\%. They are illustrated in Fig.\ \ref{fig:strucs}(f-h).

Finally, we mention some classic cases where the microhardness models have failed, and illustrate that even for these difficult systems our simple approach yields results that are sufficiently accurate to guide a CSP search. $T$-carbon (\#33, \textbf{dia-a}) is a hypothetical carbon allotrope \cite{Sheng:2011tcarbon} whose prediction prompted scrutiny of the various computational descriptors of hardness. The Gao \cite{Gao:2003a} and SV \cite{Simunek:2006a} microhardness models yielded Vickers hardnesses of 61.1~GPa and 40.5~GPa, respectfully, but it was unclear why such a porous structure with a highly anisotropic distribution of the $sp^3$-like C-C bonds should be superhard. The macroscopic Chen model yielded results that were more in-line with physical reasoning, $H_\text{v,Chen}=5.6$~GPa \cite{Chen:2011tcarbon}. The failure of the Gao and SV  models to accurately predict the hardness was attributed to the fact that they both assume uniformly distributed bonds. Herein, we find  $H_\text{v,Teter}^\text{ML}$/$H_\text{v,Teter}^\text{AEL}=8.7/7.0$~GPa for $T$-carbon, which is a reasonable estimate. 
Layered van der Waals bonded systems such as graphite have also proven to be challenging for the microhardness models. For example, the Li model \cite{Li:2008a} predicts a hardness of 57.4~GPa for graphite \cite{Lyakhov:2011a}, whereas experimental measurements yield $<1$~GPa \cite{Patterson:2000a}. The value we calculate for a graphite structure that was optimized with Grimme's D3 dispersion correction, 27.8~GPa, is too high for a quantitative estimate. However, the fitness of structures with such a small Vickers hardness would be quite low, and it is unlikely that any of them would be present in the pool in the final stages of an EA search. Herein, 62.5\% of the structures whose energy was below $-8.80$~eV/atom had an $H_\text{v,Teter}^\text{ML}$ greater than 40~GPa, confirming the preferential exploration of the PES of hard structures, which can further be tuned by adjusting the weight in the relation used to determine fitness, Eq.\ \ref{eqn:XtalOptCombinedFitnessFunction}.

\section{Conclusion}
In the past, \emph{a priori} crystal structure prediction (CSP) algorithms designed to pinpoint novel superhard phases have estimated a structure's hardness using microscopic models.  Macroscopic hardness models overcome these limitations, but they depend upon elastic properties that are expensive to calculate from first principles. 

We have shown that the Vickers hardnesses, $H_\text{v}$, of a wide range of materials computed via the Teter, Chen and Tian macroscopic models along with the bulk and/or shear moduli calculated using AFLOW-AEL agree well with experiment. Moreover, the $H_\text{v}$ calculated using the moduli obtained via a ML model that was trained on the AFLOW repository are in excellent agreement with the AFLOW-AEL results. Because the ML based values can be computed quickly given only a crystal's structure, it is possible to employ them to determine fitness in a CSP search. This methodology has been implemented in the \textsc{XtalOpt} evolutionary algorithm, which can now calculate an individual's probability for procreation using both the Vickers hardness and the energy/enthalpy, to accelerate the search for hard and stable species.

The method was applied towards the carbon system. In addition to finding 36 known phases, 43 dynamically stable novel superhard individuals were identified. Their structures were analyzed using a topological analysis, which showed that all but one of the newly predicted phases with $H_\text{v,Teter}^\text{ML}>70$~GPa were comprised of a substantial fraction of diamond and/or lonsdaleite. The other superhard phase was close to amorphous or disordered. We expect that in the future CSP algorithms will become increasingly coupled with ML derived values to predict materials with desired properties.

\section{Methods}
All of the calculations were carried out using density functional theory as implemented within the Vienna \emph{ab initio} Simulation Package (VASP) \cite{Kresse:1993a} coupled with the Perdew, Burke and Ernzerhof (PBE) functional \cite{perdew1996generalized}, and the projector augmented wave (PAW) method \cite{blochl1994projector}. 

Detailed computational settings for the AFLOW-AEL \cite{Toher:2017a}  calculations used to determine the elastic properties are described in Ref.\ \citen{Calderon:2015a}.  A set of 4 normal and shear deformations (2 compressive and 2 expansive) are applied in each independent direction (1 to 3 depending on the structure's symmetry) to the fully relaxed structure of a given compound. The ionic positions are optimized while keeping the cell size and shape fixed, and the stress tensors for each deformed structure are calculated. The resulting stress-strain data is fitted to obtain the elastic constants in the form of the symmetric $6\times6$ elastic stiffness tensor (Voigt notation). The bulk and shear moduli are calculated from elastic constants as described in Ref.\ \citen{deJong:2015a}. 

The EA searches were carried out using \textsc{XtalOpt} version r12 \cite{Zurek:2018j} on unit cells containing 8, 12, 16 and 20 formula units per primitive cell, and a total of 5624 structures were generated. These cell sizes were chosen because they are able to predict phases with 1-6, 8, 10, 12, 16 and 20 atoms in the primitive cell, and because Refs.\ \cite{Zhang:2013a} and \cite{Lyakhov:2011a} considered primitive cells with up to 24 carbon atoms. Tests showed that the likelihood of predicting superhard individuals increased when the initial volumes of the generated crystals were constrained to be about the same as that of diamond. The first generation was constructed using the \textsc{RandSpg} method \cite{Zurek:2016h}, and duplicates were removed from the breeding pool via the \textsc{XtalComp} algorithm \cite{Zurek:2011i}. The spacegroups of the predicted superhard phases were determined using AFLOW-SYM \cite{hicks2018:aflowsym}, and their structures were analyzed via the tiling approach as implemented in the ToposPro package \cite{Blatov}.

The carbon $2s^22p^2$ electrons were treated as valence, and the core states were described using the PAW method with an energy cutoff of 500~eV in the structure searches, and 600~eV otherwise.  $\Gamma$-centered Monkhorst-Pack $k$-meshes were employed where  the number of divisions along each reciprocal lattice vector was chosen such that the product of this number with the real lattice constant was 25 \AA{} for the final step in the EA searches, 40~\AA{} for the phonon calculations, and 50~\AA{} for the final optimizations and equation of state calculations. Crystals that satisfied the hardness and stability criteria described in Sec.\ \ref{sec:smallcarbon} were optimized so that all forces were smaller than 10$^{-5}$ eV/\AA.  Phonon calculations were carried out using AFLOW-APL (Automatic Phonon Library) in conjunction with VASP \cite{Curtarolo:2012a}. APL uses the finite displacement method to calculate the phonon dispersion. Supercells are generated with displaced atoms, and the interatomic forces are calculated using VASP. APL uses these forces to calculate the interatomic force constants and generate the dynamical matrix, which is diagonalized to obtain the phonon eigenmodes and dispersions.\\[2ex]


\noindent\textbf{Data Availability} 
The datasets generated during and/or analyzed during the current study are summarized in the supplementary information, and are available from the corresponding author on reasonable request. \\[2ex]

\noindent\textbf{Acknowledgements}
We acknowledge the DOD-ONR (N00014-16-1-2583 and N00014-17-1-2090) for financial support, and the Center for Computational Research (CCR) at SUNY Buffalo for computational support. \\[2ex]

\noindent\textbf{Author Contributions}
E.Z.\ conceived the research. P.A.\ and C.T.\ tested the macroscopic hardness models. P.A.\ carried out the method development within the \textsc{XtalOpt} code. P.A.\ and X.W.\ performed the evolutionary searches, used the AFLOW-ML API to predict hardness values with assistance from E.G., and analyzed the results. X.W.\ ran the phonon calculations to check dynamical stability, with assistance from C.O. D.M.P.\ led the topological analysis. E.Z.\ and S.C.\ supervised the study. All authors participated in discussing the results, and commented on the manuscript. \\[2ex]

\noindent\textbf{Additional Information} \\
\textbf{Supplementary information} accompanies the paper on the \emph{npj Computational Materials} website ().

\noindent\textbf{Competing Interests:} The authors declare no competing interests.


\end{document}